\documentclass[3p,twocolumn,times]{elsarticle}

\usepackage{framed,multirow}

\usepackage{amssymb}
\usepackage{latexsym}

\usepackage{amsmath} 
\usepackage{booktabs}
\usepackage{multirow}

\usepackage{url}
\usepackage{xcolor}

\usepackage{hyperref}

\let\today\relax
\makeatletter
\def\ps@pprintTitle{%
    \let\@oddhead\@empty
    \let\@evenhead\@empty
    \def\@oddfoot{\footnotesize\itshape
         {} \hfill\today}%
    \let\@evenfoot\@oddfoot
    }
\makeatother

\author[1]{Alessia Gerbasi\corref{cor1}}
\cortext[cor1]{Corresponding author: alessia.gerbasi01@universitadipavia.it}
\author[1]{Arianna Dagliati}
\author[1]{Giuseppe Albi}
\author[2]{Mattia Chiesa}
\author[2,3]{Daniele Andreini}
\author[2,4]{Andrea Baggiano}
\author[2]{Saima Mushtaq}
\author[2,3]{Gianluca Pontone}
\author[1,6]{Riccardo Bellazzi\corref{cor2}}
\author[2]{Gualtiero Colombo\corref{cor2}}
\cortext[cor2]{Co-last authors}
\address[1]{Department of Electrical, Computer and Biomedical Engineering, University of Pavia, Pavia, Italy}
\address[2]{Centro Cardiologico Monzino IRCCS, Milan, Italy}
\address[3]{Department of Biomedical and Clinical Sciences, University of Milan, Milan, Italy}
\address[4]{Department of Clinical Sciences and Community Health, University of Milan, Milan, Italy}
\address[5]{Department of Biomedical, Surgical and Dental Sciences, University of Milan, Milan, Italy}
\address[6]{IRCCS Istituti Clinici Scientifici Maugeri, Pavia, Italy}

\begin{document}

\begin{frontmatter}

\title{CAD-RADS scoring of coronary CT angiography with Multi-Axis Vision Transformer: a clinically-inspired deep learning pipeline}

\begin{abstract}

The standard non-invasive imaging technique used to assess the severity and extent of Coronary Artery Disease (CAD) is Coronary Computed Tomography Angiography (CCTA). However, manual grading of each patient's CCTA according to the CAD-Reporting and Data System (CAD-RADS) scoring is time-consuming and operator-dependent, especially in borderline cases.\\ 
This work proposes a fully automated, and visually explainable, deep learning pipeline to be used as a decision support system for the CAD screening procedure. The pipeline performs two classification tasks: firstly, identifying patients who require further clinical investigations and secondly, classifying patients into subgroups based on the degree of stenosis, according to commonly used CAD-RADS thresholds. 
The pipeline pre-processes multiplanar projections of the coronary arteries, extracted from the original CCTAs, and classifies them using a fine-tuned Multi-Axis Vision Transformer architecture. With the aim of emulating the current clinical practice, the model is trained to assign a per-patient score by stacking the bi-dimensional longitudinal cross-sections of the three main coronary arteries along channel dimension. Furthermore, it generates visually interpretable maps to assess the reliability of the predictions. When run on a database of 1873 three-channel images of 253 patients collected at the Monzino Cardiology Center in Milan, the pipeline obtained an AUC of 0.87 and 0.93 for the two classification tasks, respectively.\\
According to our knowledge, this is the first model trained to assign CAD-RADS scores learning solely from patient scores and not requiring finer imaging annotation steps that are not part of the clinical routine.

\end{abstract}

\end{frontmatter}

\section{Introduction}
\label{sec:intro}
Coronary artery disease (CAD) is the leading cause of cardiovascular mortality worldwide. It is caused by \textit{atherosclerosis}, a phenomenon consisting in the formation of plaques that gradually narrow the diameter of the arteries reducing the oxygen-rich blood supply to the heart. The disruption of these plaques is the main cause of the acute coronary syndromes (i.e., unstable angina, myocardial infarction) \citep{falk1995coronary}. Although the underlying causes for the development of atherosclerosis are not yet fully understood, there are numerous risk factors that can lead to accelerated plaque formation, such as high low-density lipoprotein cholesterol (LDL) level, high blood pressure, diabetes mellitus, smoking, obesity, advancing age and family predisposition. For these reasons, identification and monitoring of patients at high risk of CAD through noninvasive procedures are of utmost importance.

Cardiac computed tomography angiography (CCTA) allows noninvasive identification of coronary stenosis and high-risk plaque features, which is useful for risk stratification. Recently, CCTA has been integrated into the routine clinical management of patients with suspected CAD because of its value as an effective rule-out tool.
Imaging software included in the most commonly used CT scanners, usually include multiplanar reconstruction (MPR) and straightened curved planar reformation (CPR) methods \citep{kanitsar2002cpr} able to trace the blood vessels and generate 2D longitudinal cross-sections from the CCTA scans. These representations are extremely useful to better visualize the vessel of interest without surrounding structures that could make it harder for the physician to identify the plaque from the original three-dimensional scan. 

With the aim of creating a standardized method to communicate imaging findings, CAD-Reporting And Data System (CAD-RADS) scoring has been proposed and it is currently used in the clinical practice \citep{cury2016cad}. According to CAD-RADS classification system, each patient can be classified with a score ranging from 0 to 5. A score of 0 indicates absence of CAD; 1 corresponds to stenosis between 1-24\%; 2 to stenosis between 25-49\%; 3 to stenosis between 50-69\%; 4 to stenosis between 70-99\% or $>$50\% left main or three vessels $>$70\%; 5 to total occlusion. One of the main limitations associated with manual scoring of CCTA scans is its dependence on the physician expertise, which can be crucial in borderline cases. On the other hand, automating this process is challenging because the CAD-RADS is a per-patient score and it is assigned on the basis of a visual assessment of the degree of occlusion of the three major coronary arteries: left anterior descending artery (LAD), left circumflex artery (LCX) and right coronary artery (RCA). 

In the last years, deep learning (DL) models have been widely explored in the medical field. If correctly designed and evaluated, these methods offer a chance to enhance healthcare accessibility, fairness, precision, and inclusivity \citep{esteva2021deep}. In particular, convolutional neural networks (CNNs) have dominated the field of computer vision in the past years and are still the most widely used models for solving tasks ranging from segmentation to classification. By employing filters, these networks are able to learn feature maps that highlight the most relevant parts of the input images. Visual transformers have recently gained a great popularity in computer vision achieving state-of-the-art (SOTA) performance in many visual tasks \citep{dosovitskiy2020image}. The main advantage of transformer architectures is their attention mechanism. However, when compared to classical CNNs, their reduced inductive bias can easily lead to overfitting. This is the reason why their superiority to convolutional models is generally appreciable when large data sets are available. In the medical context, this is almost never the case since several factors ranging from data privacy to heterogeneity and lack of standard quality, makes it difficult to create large and consistent datasets. Therefore, simpler models to mitigate the risk of overfitting and to facilitate output interpretation are usually preferred. This can often lead to sub-optimal results, given the well-known limitations of many of the most popular convolutional models compared to the most recently proposed ones.
Many improvements of standard convolutions have been recently proposed to make them more efficient (e.g. \cite{howard2017mobilenets, sandler2018mobilenetv2}). On the other hand, many recent works have tried to improve scalability of attention mechanisms (eg.\cite{liu2021swin}), or to propose hybrid methods such as \cite{dai2021coatnet,xiao2021early}.
In particular, \cite{tu2022maxvit} recently proposed MaxViT, a hybrid model combining the strengths of both approaches (efficient convolutions and sparse attention) in a new “base-block” able to significantly improve upon SOTA performance \textit{under all data regimes} for many visual tasks, including image classification. This new base-block consists of a MB-Conv block \citep{howard2017mobilenets} with a squeeze-and-excitation (SE) module \citep{hu2018squeeze} followed by a multi-axis attention block appositely designed to capture both local and global pixels interactions.

In the light of these considerations, this paper proposes a DL pipeline for the automatic classification of straightened MPR images obtained from CCTA scans, based on MaxVit architecture.
The goal of the study is to automate the CAD-RADS scoring process with an explainable decision support system able to (1) rule-out patients needing for further investigations and (2) classify the patients into three main groups based on the degree of stenosis, according to commonly used CAD-RADS thresholds.
Our aim is to build a fully automated DL pipeline able to guide the physician in the clinical practice, providing a tool which is at the same time accurate and easy to interpret by the final user. 

The main contributions of this study are:
\begin{itemize}
    \item The development of a novel fully automated pipeline based on MaxViT architecture specifically trained to assign a patient-based CAD-RADS score. As far as we know this is the first approach tackling the CAD-RADS scoring problem emulating the clinical procedure.
    \item The design of a flexible approach not requiring vessel, segment or lesion-wise annotations and considering the three main coronary arteries. 
    \item An extensive experimentation on a curated dataset of 253 patients presenting quantitative evaluations and visually explainable results.
\end{itemize}

\section{Related works}
\label{sec:relworks}
Several different approaches have been proposed with the aim of automating the identification and grading of coronary stenosis. We report the most recently proposed works that exploit DL methods. 

\cite{huang2022clinical} showed that there is no significant difference between the DL-based (convolutional models in this case) and the expert-based CAD-RADS grading of CCTAs (Kappa value of 0.77). This result is very interesting from a clinical perspective because it suggests the high potentiality of DL based decision support systems for this particular clinical tasks.

\cite{li2022automatic} developed a coronary tree segmentation algorithm (Dice score 0.771) and proposed a binary classification algorithm (3DNet) taking as input the segmented tree and other relevant clinical features, with the aim of predicting patient-wise CAD-RADS score achieving a diagnostic performance in terms of area under the ROC curve of 0.737.

\cite{denzinger2020automatic} proposed a DL strategy that reaches a ROC AUC of 0.923 on the task of identifying patients with a CAD-RADS score $>2$ that was then improved in a more recently proposed version to 0.950 \citep{denzinger2022cad}. The proposed method is very promising and tested on a large cohort of patients, but requires segment-level annotations, a step that is not usually part of clinical routine. 

Other works focused instead on single lesion or single vessel scoring automated systems. However, deriving patient scores based on individual lesions can lead to a significant number of potential errors, as it fails to take into account the overall context in making decisions. 
\cite{paul2022evaluation} for example, achieved a 96\% accuracy in identifying significant stenosis from a huge dataset of curved multiplanar reformatted (cMPR) CCTA images originally classified by an expert radiologist. However, the single vessel grading strategy is time consuming, highly influenced by the radiologist expertise and not usually part of clinical routine.

\cite{penso2023token} proposed a token-mixer architecture for CAD-RADS classification achieving 82\% of accuracy in classifying significant stenosis and 72\% in a multi-class experimental set-up predicting CAD-RADS 0 vs. 1–2 vs. 3–4 vs. 5. Even in this case, each coronary artery was individually labelled.

\cite{tejero2019texture} proposed a model leveraging multiple feature extractors for texture classification using multiple CPR views of the coronary arteries. The method shows good performance in predicting significant stenosis on a dataset of 57 patients. Although the limited dataset size and the need for manual ground-truth annotation of the stenosis, they achieved $80\%$ of accuracy using a leave-one-out cross-validation strategy. 

\cite{candemir2020automated} proposed a 3D-CNN obtaining good performance for coronary artery atherosclerosis detection on MPR volumes. The algorithm uses pre-processing techniques and a 3D-CNN to identify atherosclerotic plaques and provides visual clues for location. The method obtains an accuracy of $90.9\%$ in identifying patients with atherosclerosis. The authors proposed it as a method for assisting physicians in excluding coronary atherosclerosis in patients with acute chest pain. 

\cite{muscogiuri2020performance} demonstrated how DL methods for CAD-RADS scoring are significantly faster if compared to human on site reading of clinical scans. They proposed several small custom 2D-CNN models with the best one achieving an accuracy of 81\% in classifying patients with CAD-RADS 0 vs. CAD-RADS $>$0. The models were trained using single 2D slices from original CCTA scans without extracting the coronary arteries.

A recurrent CNN was proposed by \cite{zreik2018recurrent} for automatic detection and classification of coronary artery plaque and stenosis, achieving an accuracy of 0.77 and 0.80, respectively, on a test set of 65 patients. In this last work the presence and the anatomical significance of coronary stenosis were manually annotated on the MPR images. 

Most of the discussed works present interesting different approaches from both technical and clinical perspective, however they all require vessel, segment, lesion annotations or derive patient-scores considering single lesions. While these approaches include more information if compared to a single patient-wise score, they require an additional effort in the clinical practice since segment or lesion-wise annotation is not a routine operation. Moreover, they are not able to take into account the global patient status. 

Differently from these methods, our aim is to simulate the real clinical approach, where an expert physician visually inspects multiple MPR views of the three main coronary arteries and, based on his experience and on CAD-RADS guidelines, assigns a patient-level score, bypassing finer annotation steps. To reach our goal, we developed a novel pipeline based on a recently proposed neural architecture, which is able to accurately emulate the clinical scoring procedure.

\begin{figure*}[ht]
    \centering
    \includegraphics[width=0.99\textwidth]{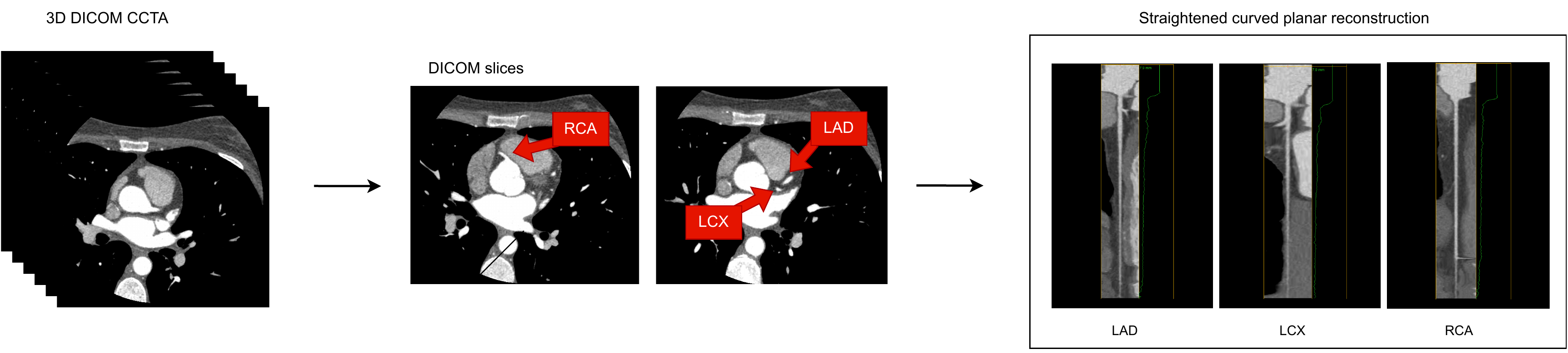}
    \caption{Example of CCTA scan for a random patient included in the study population. From the 3D DICOM scan we extracted two slices where the three main coronary arteries are indicated in red. On the right we can see the 2D representation of LAD, LCX and RCA obtained through straightened curved planar reconstruction.}
    \label{fig:ccta}
\end{figure*}

\section{Materials and Methods}
\label{sec:methods}
\begin{figure*}[ht!]
    \centering
    \includegraphics[width=0.89\textwidth]{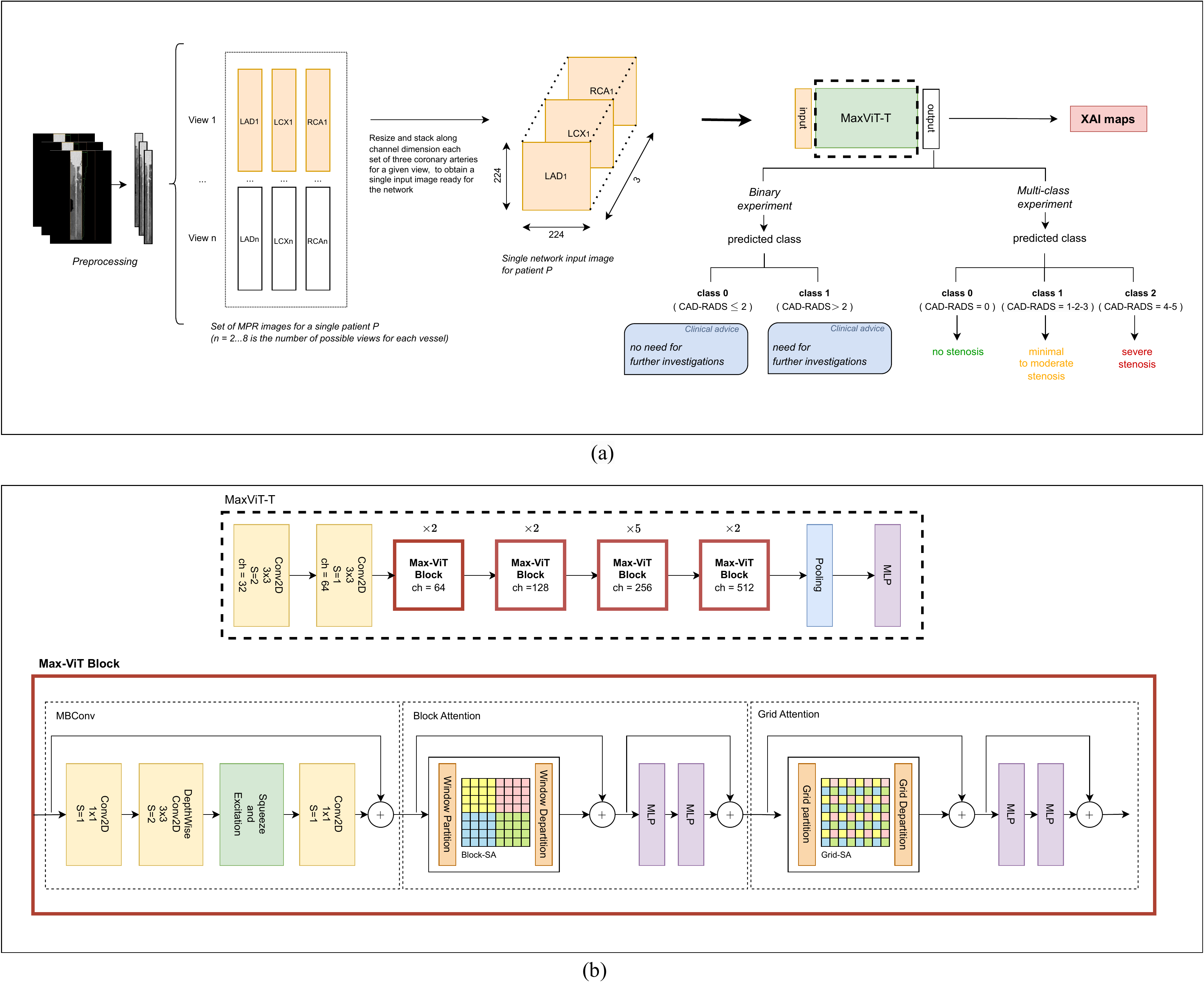}
    \caption{(a) Schematic representation of the main pipeline steps. All the MPR projections are first of all pre-processed to enhance the image contrast. Afterwards, for each patient, an imputing step is performed in order to always have three 2D images representing RCA, LAD and LCX for each of the $n$ different views (where $n$ = 2...8 is the number of possible views for each vessel). For each view, the 3 images are then resized (224x224) and stack along channel dimension in order to obtain a single input image ready for the network. The images thus created are used to fine-tune a MaxViT-T architecture to solve two different tasks (binary and multi-class). Finally, SOTA eXplainable AI (XAI) models are used to create qualitative maps to visually inspect the reliability of network's predictions.
    (b) Max-ViT-Tiny architecture used in the proposed pipeline. It is composed by two convolutional blocks, followed by several MaxViT blocks and a final pooling layer that precedes the MLP head. The architecture of each MaxViT block is schematized in the figure: there is an initial MBCov block followed by a block-attention and a grid-attention block.}
    \label{fig:pipeline}
\end{figure*}

We set up two different experiments following the same analysis pipeline. In the first one (named \textit{binary experiment}) we binarized the CAD-RADS score with a threshold of 2 (0-1-2 vs. 3-4-5) with the aim of simply distinguishing patients in need for further examinations or direct intervention (CAD-RADS $>$ 2). In the second experiment (named \textit{multi-class experiment}) we trained the model to predict 3 different classes: healthy subjects (CAD-RADS = 0), patients with minimal to moderate stenosis (CAD-RADS = 1-2-3) and patients with severe stenosis or complete occlusion (CAD-RADS = 4-5). This second approach would be useful to quickly identify completely healthy subjects as well as patients with very severe stenosis, grouping the intermediate or borderline cases that could need a more accurate inspection from the physicians. The complete pipeline is illustrated in Figure \ref{fig:pipeline}, each step is described in the following subsections.

\subsection{Dataset}
\label{subsec:data}
We applied our novel pipeline to a dataset of $253$ patients who underwent CCTA for clinical purposes from 2016 to 2018 in the Monzino Cardiology Center (Milan, Italy). The original population is fully described by \cite{muscogiuri2020performance}. Exclusion criteria for this study were heart rate $\ge80$ bpm despite intravenous administration of beta blockers, atrial fibrillation, BMI $\ge35$ kg/m$^2$ \citep{pontone2018impact} and presence of stent. Sublingual nitrates were administered 5 minutes before the CCTA scan \citep{takx2015sublingual}. 
Two CT scanners, the Discovery CT 750 HD and Revolution CT (GE Healthcare, Milwaukee, IL), were used for CCTA acquisition. 
The CCTA protocol defines a $64\times0.625$ mm and a $256\times0.625$ mm slice configuration for the Discovery CT 750 HD and the Revolution CT, respectively. The tube current and voltage where adjusted based on the patient's BMI \citep{pontone2012feasibility}.
In both protocols, 50–70 mL of contrast medium was given through the antecubital vein at an infusion rate of 5 mL/s, followed by 50 mL of saline solution at the same rate. The bolus tracking technique was used for CCTA acquisition, and images were reconstructed using filtered back projection and in 75\% or 40–80\% of the cardiac cycle, depending on the ECG-triggering acquisition used \citep{pontone2018image}. 
In cases of poor image quality, intracycle motion correction was performed \citep{pontone2016impact,pontone2018image}.
A consensus of five different random couples between ten radiologists and cardiologists was formed to score the pool of CCTA examinations. The cardiac imagers had experience ranging from 5 to 10 years. A CAD-RADS score was attributed for each examination, and in cases of disagreement, a cardiac imager with 10 years of experience in cardiovascular imaging adjudicated the final CAD-RADS score.

For each patient and each coronary artery (RCA, LCX, LAD) up to eight straightened MPR views were extracted from the original CCTA, with a 45° angle offset. In particular, if the subject was classified with CAD-RADS = 0 (0\% stenosis), we always had exactly 8 images for each main coronary artery in our dataset. On the other hand, for patients with CAD-RADS $>$ 0, only images from non-healthy coronaries were collected
(eg. if a patient has a CAD-RADS score of 3 and only the LAD and LCX arteries present with stenosis, we have up to 8 views for each of these two vessels, but no images for the healthy RCA). Therefore, in our dataset, completely healthy coronary arteries, which do not influence the CAD-RADS score when this is greater than 0, were discarded a priori by the clinicians at the data acquisition stage. 
Our aim is to fully automate the process and train an algorithm to predict the CAD-RADS score from the three main coronary arteries without any prior knowledge. For this reason the missing healthy vessels were imputed before the classification step as fully described in Section \ref{subsec:imputing}.
In Figure \ref{fig:ccta} we can see an example CCTA for a patient included in our dataset. We show two slices of the original 3D DICOM scan where it is possible to notice the 3 main coronary arteries in a bi-dimensional space and, on the right, we can see the LAD, LCX and RCA images resulting from the straightened curved planar reconstruction.

\subsection{Preprocessing}
\begin{figure}[ht]
    \centering
    \includegraphics[scale=0.8]{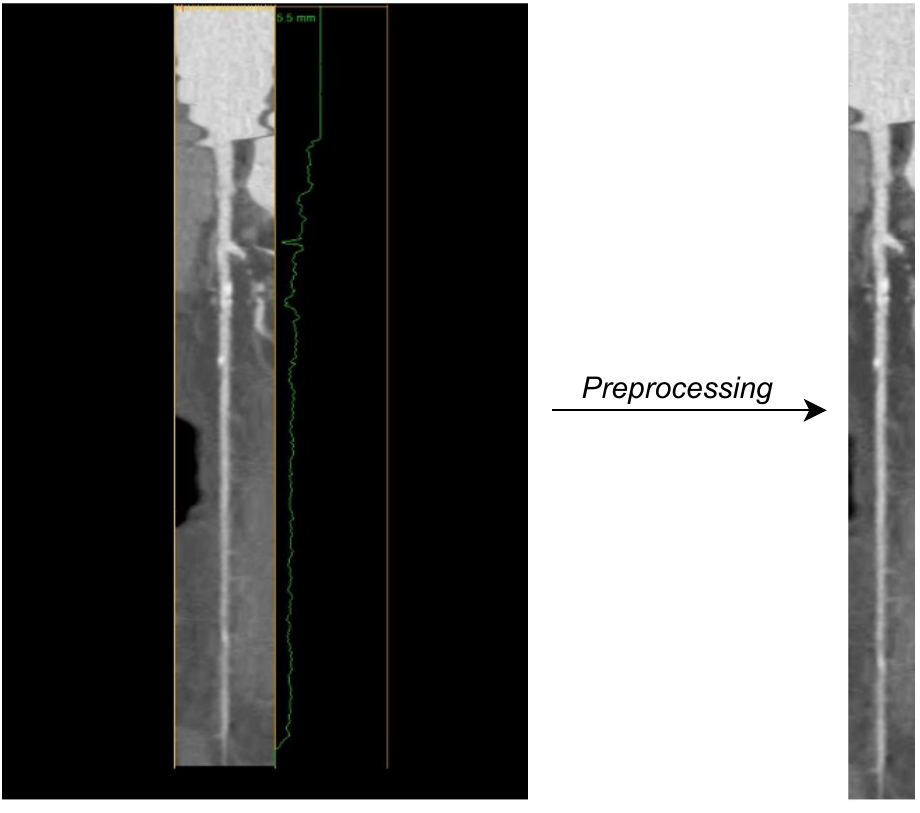}
    \caption{Sample image before and after the preprocessing steps:  (1) Annotation/artefacts removal, (2) Contrast enhancement (CLAHE), (3) Background crop.}
    \label{fig:preprocessing}
\end{figure}

The first step of our pipeline is image preprocessing. Our input data is composed by 2D images representing different views of the straightened MPR volume for each patient. In the first step we removed artefacts on digital scans by binarizing each image, sorting white objects by size and keeping just the largest one (representing the vessel). Therefore, we made sure to delete all annotations and small artefacts derived from image reconstruction. Finally, we applied Contrast Limited Adaptive Histogram Equalization (CLAHE) to enhance the local contrast of the image \cite{pizer1987adaptive}. As a final step we automatically cropped the images to reduce background black pixels on the 4 sides. A sample image before and after the preprocessing steps is showed in Figure \ref{fig:preprocessing}.

\subsection{Dataset split and imputing}
\label{subsec:imputing}
After the preprocessing, data was randomly split into training (80\%) and test (20\%) set. The training set was then further split into training and validation set using a 10-fold cross-validation strategy. All the splits were always done patient-wise and stratified by CAD-RADS score in order to avoid any selection bias. Images from completely healthy vessels for patients with CAD-RADS $>0$ were missing in our dataset as previously described in Section \ref{subsec:data}. Therefore, we averaged (separately for each view) the images of RCA, LCX and LAD coronaries of the healthy subjects (CAD-RADS = 0) included in the training set and used them to impute the missing data in our dataset. This step allowed us to always have three coronary arteries to evaluate, to simulate as closely as possible the decision-making process actually followed in clinical practice. 

\subsection{Model architecture and training strategy}
Given the limited size of our dataset we decided to fine-tune the MaxViT-T (where T stands for tiny) pretrained on ImageNet \citep{deng2009imagenet}. This recently proposed architecture combines the strengths of efficient convolutions and attention mechanism achieving SOTA classification performance under all data regimes. The architecture details are showed in Figure \ref{fig:pipeline}b. It is composed by 2 convolutional blocks followed by multiple groups of MaxViT blocks. A single MaxViT block is always composed by an initial MBConv block, also called inverted residual block, that follows a narrow $\rightarrow$ wide $\rightarrow$ narrow structure approach which greatly reduces the number of parameters if compared to a standard residual block \citep{sandler2018mobilenetv2}. As it is shown in the figure, between the $3\times3$ depthwise convolution and the final $1\times1$ convolution, there is a SE module \citep{hu2018squeeze} able to model interdependencies between channels. Following the MBConv block, we always have a block-attention and a grid-attention block which are used to capture local and global patterns respectively. Each group of MaxViT blocks differs from the other by the number of spatial filters in the convolutional layers. Finally, an average pooling layer precedes the MLP head used to classify the input into 2 (binary experiment) or 3 classes (multi-class experiment). 

To further reduce the risk of overfitting, we implemented online data augmentation on the training set with random rotation and horizontal/vertical flip, learning rate and weight decay and label smoothing. The optimal parameters were tuned with a grid search strategy based on average validation set accuracy. A complete list of the parameters used is available in Section \ref{subsec:expsetup}. 
As showed in Figure \ref{fig:pipeline}, after the preprocessing and imputing steps previously described, for each view, the 3 images (representing LAD, LCX and RCA respectively) are then resized (224x224 pixels) and stack along channel dimension in order to obtain a single input image ready for the network. This approach allows us to train the network to classify a sequence of three coronary arteries as belonging to one of the classes representing patient-wise CAD-RADS scores and thus simulating the classification process followed in clinical practice.

\subsection{Experimental setup}
\label{subsec:expsetup}
After the preprocessing steps we obtain a total number of 5619 one-channel images, and consequently 1873 three-channel images for 253 patients. Models were trained for 50 epochs using AdamW optimizer \citep{Loshchilov2019} for both the experiments, while binary cross-entropy and weighted cross-entropy loss were used for the binary and multi-class experiment, respectively. 
 
 Table \ref{tab:params} summarizes the best hyperparameters resulting from the grid-search according to the average validation accuracy during the cross-validation procedure. The tuned hyperparameters are: learning rate (LR) $\in \{1e^{-3}, 1e^{-4}, 1e^{-5}\}$, Drop-out $\in \{0.1, 0.3, 0.5\}$, weight decay (L2) $\in \{1e^{-1}, 1e^{-2}\}$, LR decay epoch $\in\{20,30\}$, label smoothing $\in\{0.1,0.2\}$. 
 
\begin{table}[t]
\caption{Set of model's hyperparameters leading to best average validation accuracy during the grid-search procedure on the training set.}
\label{tab:params}
\centering
\resizebox{\columnwidth}{!}{\begin{tabular}{@{}lccccc@{}}
\toprule
Experiment           & Lr  & Lr decay epoch & Drop-out & L2 & Label smoothing \\ \midrule
\textit{Binary}      & $1e^{-4} \rightarrow 1e^{-5}$    & $30$ &  $0.5$        &  $0.1$  &  $0.1$               \\
\textit{Multi-class} &  $1e^{-4} \rightarrow 1e^{-5}$  & $30$ &    $0.3$      & $0.01$  & $0.2$                \\ \bottomrule
\end{tabular}}
\end{table}

The best hyperparameters are then used to train the models on the whole training set and evaluate the performance on the test set.

The performance of the chosen architecture was also compared with several other fully convolutional (ResNet18, ResNet50 \citep{he2016deep}; Vgg16, Vgg19 \citep{simonyan2014very}), attention-based (ViT-T \citep{dosovitskiy2020image}) or hybrid models (ConvNeXt-T \citep{liu2022convnet}, CoAtNet-T \citep{dai2021coatnet}). All the results are reported in Section \ref{sec:results}.

\subsection{Hardware and software}
All the models were trained using an NVIDIA 3070 Ti GPU. The whole pipeline is developed in Python 3.10. Pytorch \citep{paszke2019pytorch} is the framework used for implementing the models. Open-CV implementation of CLAHE algorithm was used for enhancing the contrast in the preprocessing step \citep{opencv} and DeepSHAP \citep{NIPS2017_7062} library for the final explainability maps.

\subsection{Visual interpretation}
One of the main drawbacks of DL models is their difficult interpretability, which has been tackled with explainability approaches \citep{gunning2019xai,xu2019explainable}. Furthermore, it is of foremost importance to evaluate models reliability and assure systems trust \citep{jones2021enhancing}, especially in the medical field, where final users might not be aware of the methodological implementation of the supporting system. The reliability of a decision support system strongly increases when it becomes easily interpretable by the final user, who, based on his experience, could simply detect edge cases in which the algorithm results may not be trustable. With this aim in mind, we added three levels of visual explainability of the results:
\begin{itemize}
    \item \textbf{t-SNE} \citep{van2008visualizing} plot to project the last layer features down to a 2D space in order to evaluate the semantic understanding of the network.
    \item \textbf{maximally activated patches} \citep{zeiler2014visualizing} representing the most responsive areas of an image, which are identified through a forward pass with that image that is partially “occluded”. This is accomplished by masking portions of the image (patches) and evaluating the impact on the predicted scores for the top class. The patches that cause the greatest change in the scores are prioritized and the top-k of these patches are then visualized. 
    \item \textbf{Deep SHAP} \citep{NIPS2017_7062} explainability maps. This algorithm is a fast approximation method for computing SHAP (SHapley Additive exPlanations) values in DL models. It is a game theoretic approach that generates visual maps for each image where pink pixels indicate the image values that contributed to the model's prediction of a specific output class, while blue pixels represent the values that pushed the prediction towards the alternative class. This visual representation allows for inspection of pixels that were most significant in determining the final classification, according to the model. This approach is particularly interesting since it has been shown to align better with human intuition compared to other explainability methods.
\end{itemize}

\section{Results}
\label{sec:results}
 Final results on the test set for both experiments are provided in Table \ref{tab:res}. Algorithms performance was measured in terms of area under the ROC curve (AUC), accuracy, precision, recall, F1 score. All the results are presented both image-level and patient-level. In the second case, we considered all the images from the same patient (representing different views of the same vessels) and we assigned as final class the one with higher average predicted probability.

\begin{table*}[ht]
\caption{Results of the \textit{binary} and \textit{multi-class experiment} computed on the test set. For all the metrics 95\% confidence interval is provided.}
\label{tab:res}
\resizebox{\textwidth}{!}{\begin{tabular}{lcccccccc}
\hline
\multicolumn{1}{c}{\textbf{Experiment}} & \textbf{Class}                & \textbf{Type of metric} & \multicolumn{1}{c}{\textbf{AUC [95\% CI]}}      & \multicolumn{1}{c}{\textbf{Accuracy [95\% CI]}} & \textbf{Precison [95\% CI]} & \textbf{Recall [95\% CI]} & \textbf{F1-score [95\% CI]} & \textbf{n}           \\ \hline
                                        & \multicolumn{1}{l}{}          & \multicolumn{1}{l}{}    &                                                &                                                & \multicolumn{1}{l}{}       & \multicolumn{1}{l}{}     & \multicolumn{1}{l}{}       & \multicolumn{1}{l}{} \\
\multirow{2}{*}{\textit{Binary}}                 & \multirow{2}{*}{1}            & per image               & \multicolumn{1}{c}{\textbf{0.89 [0.86, 0.93]}} & \multicolumn{1}{c}{0.82 [0.78, 0.86]}          & 0.88 [0.83, 0.93]          & 0.75 [0.68, 0.80]        & 0.81 [0.76, 0.84]          & 374                  \\
                                        &                               & per patient             & \multicolumn{1}{c}{\textbf{0.87 [0.76, 0.95]}} & \multicolumn{1}{c}{0.82 [0.72, 0.93]}          & 0.89 [0.75, 1.00]          & 0.71 [0.55, 0.89]        & 0.79 [0.69, 0.90]          & 51                   \\
                                        & \multicolumn{1}{l}{}          & \multicolumn{1}{l}{}    &                                                &                                                & \multicolumn{1}{l}{}       & \multicolumn{1}{l}{}     & \multicolumn{1}{l}{}       & \multicolumn{1}{l}{} \\ \hline\\
\multirow{11}{*}{\textit{Multi-class}}           & \multirow{2}{*}{0}            & per image               & \multicolumn{1}{c}{0.93 [0.84, 0.96]}          & \multicolumn{1}{c}{0.94 [0.92, 0.96]}          & 0.90 [0.83, 0.97]          & 0.82 [0.75, 0.90]        & 0.86 [0.82, 0.90]          & 80                   \\
                                        &                               & per patient             & 0.94 [0.81, 0.99]                              & 0.96 [0.91, 0.10]                               & 1.00 [1.00, 1.00]          & 0.80 [0.60, 1.00]        & 0.89 [0.80, 0.98]          & 10                   \\
                                        & \multicolumn{1}{l}{}          & \multicolumn{1}{l}{}    &                                                &                                                & \multicolumn{1}{l}{}       & \multicolumn{1}{l}{}     & \multicolumn{1}{l}{}       & \multicolumn{1}{l}{} \\
                                        & \multirow{2}{*}{1}            & per image               & 0.87 [0.84, 0.95]                              & 0.82 [0.78, 0.86]                              & 0.79 [0.73, 0.85]          & 0.86 [0.81, 0.91]        & 0.82 [0.78, 0.86]          & 182                  \\
                                        &                               & per patient             & 0.91 [0.83, 0.98]                              & 0.84 [0.74, 0.94]                              & 0.81 [0.67, 0.95]          & 0.93 [0.83, 1.00]        & 0.87 [0.78, 0.96]          & 27                   \\
                                        & \multicolumn{1}{l}{}          & \multicolumn{1}{l}{}    &                                                &                                                & \multicolumn{1}{l}{}       & \multicolumn{1}{l}{}     & \multicolumn{1}{l}{}       & \multicolumn{1}{l}{} \\
                                        & \multirow{2}{*}{2}            & per image               & 0.91 [0.84, 0.96]                              & 0.87 [0.84, 0.90]                               & 0.82 [0.75, 0.89]          & 0.75 [0.67, 0.83]        & 0.78 [0.74, 0.82]          & 112                  \\
                                        &                               & per patient             & 0.93 [0.84, 0.99]                              & 0.88 [0.79, 0.97]                              & 0.83 [0.62, 1.00]          & 0.72 [0.51, 0.95]        & 0.77 [0.68, 0.97]          & 14                   \\
                                        & \multicolumn{1}{l}{}          & \multicolumn{1}{l}{}    &                                                &                                                & \multicolumn{1}{l}{}       & \multicolumn{1}{l}{}     & \multicolumn{1}{l}{}       & \multicolumn{1}{l}{} \\
                                        & \multirow{2}{*}{Weighted avg} & per image               & \textbf{0.90 [0.86, 0.92]}                     & 0.86 [0.83, 0.89]                              & 0.82 [0.78, 0.86]          & 0.82 [0.78, 0.86]        & 0.82 [0.78, 0.86]          & 374                  \\
                                        &                               & per patient             & \textbf{0.93 [0.84, 0.99]}                     & 0.88 [0.79, 0.95]                              & 0.85 [0.76, 0.93]          & 0.84 [0.74, 0.92]        & 0.84 [0.73, 0.92]          & 51                   \\
                                        & \multicolumn{1}{l}{}          & \multicolumn{1}{l}{}    &                                                &                                                & \multicolumn{1}{l}{}       & \multicolumn{1}{l}{}     & \multicolumn{1}{l}{}       & \multicolumn{1}{l}{} \\ \hline
\end{tabular}}
\end{table*}

\begin{figure}[ht]
    \centering
    \includegraphics[scale=0.37]{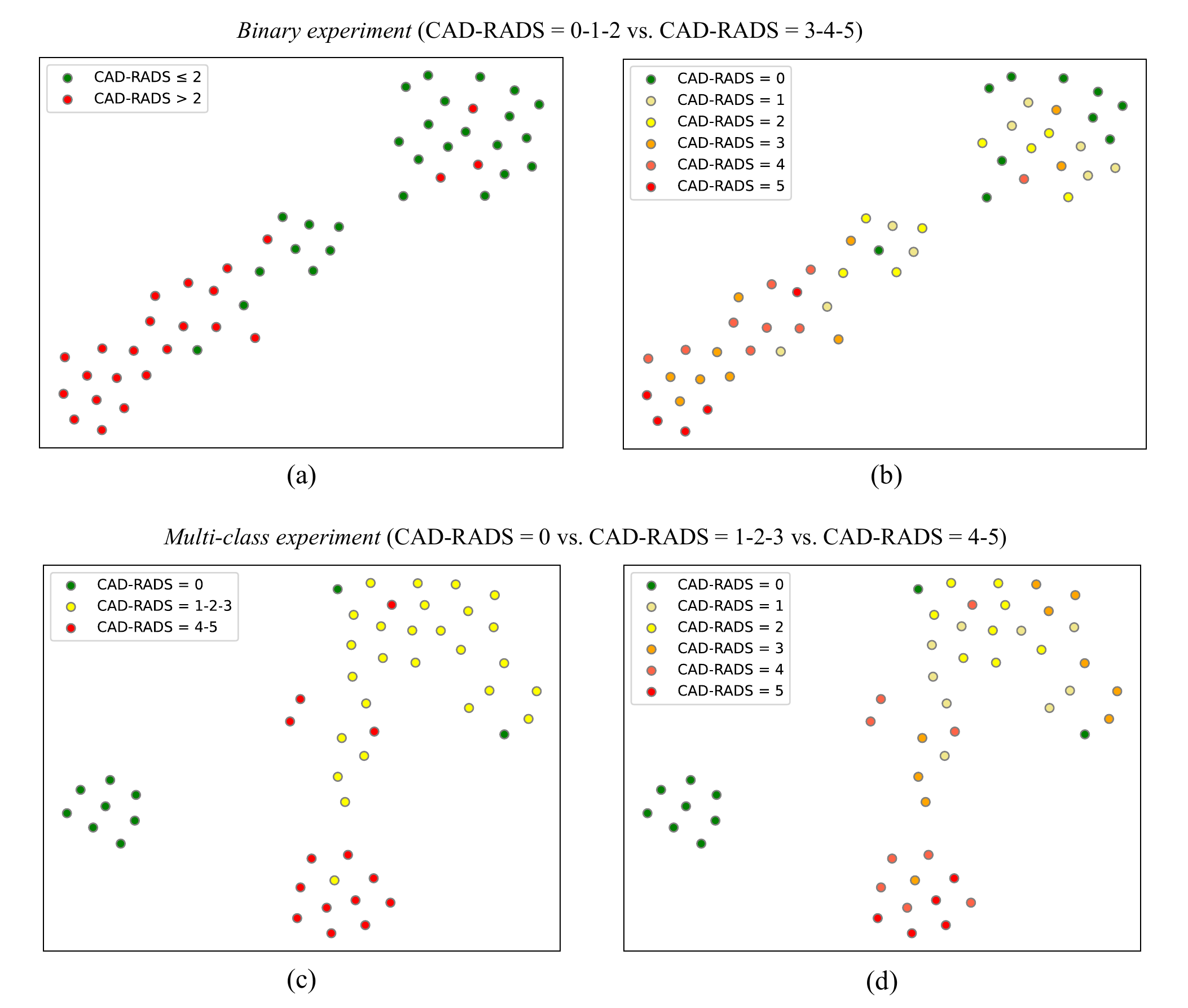}
    \caption{t-SNE plots showing a 2D representation of the 256 features representing the average patient embeddings. Results for the binary and multi-class experiment are reported in the first (a,b) and second (c,d) row, respectively. Each dot represents a patient of the test set. On the left side we colored the dots based on the labels used for training the network (a,c), on the right side instead, they are colored according to the original CAD-RADS classification (b,d).}
    \label{fig:tsne}
\end{figure}

In Figure \ref{fig:tsne} we reported per patient t-SNE plots of the binary and the multi-class experiment in the first (a,b) and second row (c,d), respectively. The dots are colored by the labels used by the algorithm on the left (a,c) and the original CAD-RADS scores on the right (b,d), to visually inspect the extent of the errors. 

\begin{figure*}[ht!]
    \centering
    \includegraphics[scale=0.5]{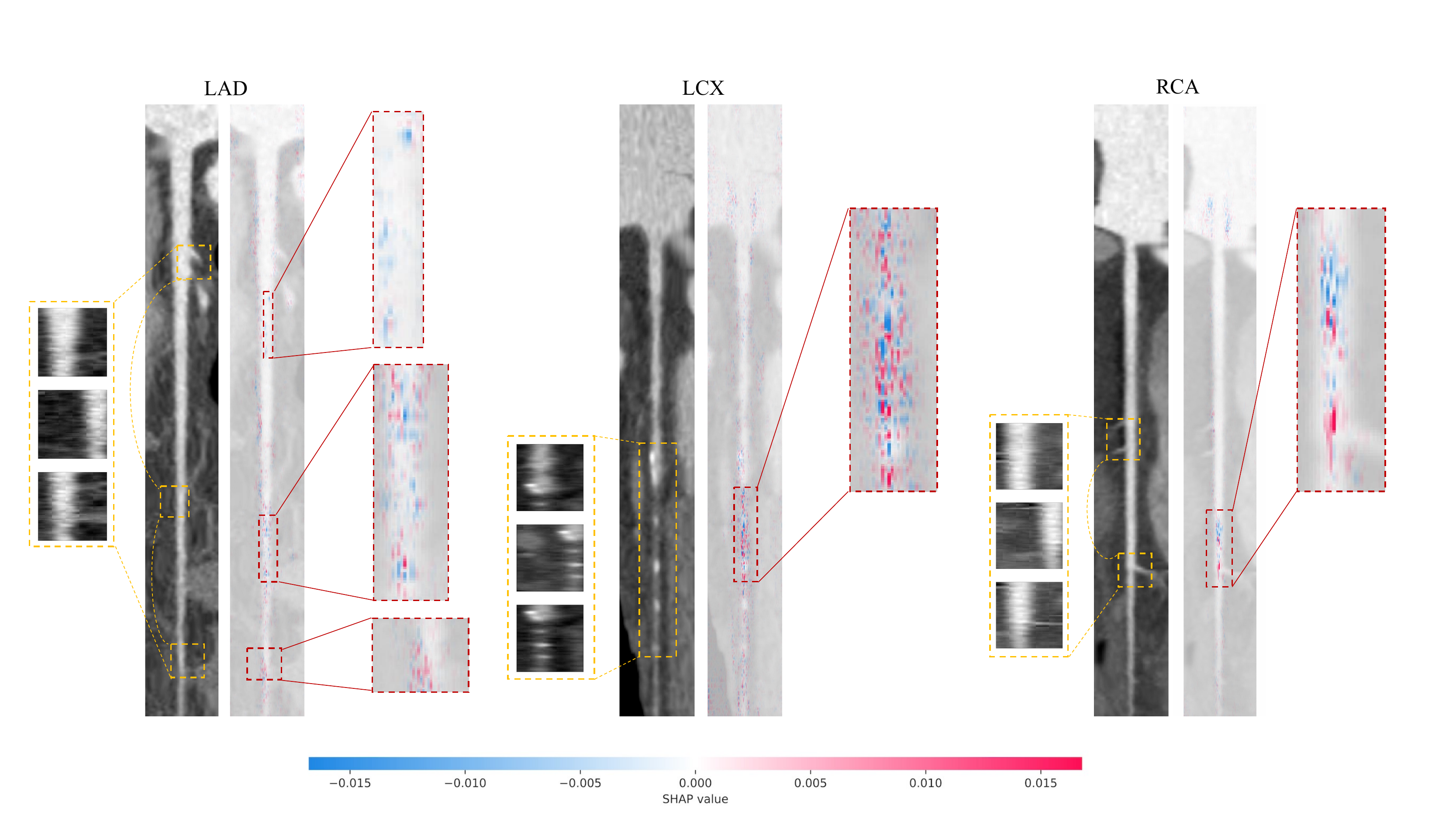}
    \caption{The figure shows the three main coronary arteries of a patient from the test set with a CAD-RADS score of 5 and correctly classified as belonging to class 2 (CAD-RADS = 4-5) by the multi-class algorithm. The three vessels (LAD, LCX, RCA) represent the three channels of the input tensor pre-processed by the network as a single input image. For each vessel, we can see the original pre-processed scan with the top-3 maximally activated patches in order of importance on the left and the Deep SHAP map on the right, with a zoom on the most relevant part of the images. Deep SHAP maps are reported for the correctly predicted output class (class 2). The input images are shown as nearly transparent grayscale backings behind each of the explanations. Pink pixels increase the model's output while blue pixels decrease the output.}
    \label{fig:shap}
\end{figure*}

Finally, Figure \ref{fig:shap} shows an example of maximally activated patches and Deep SHAP maps generated for a random patient from the test set of the multi-class experiment. The selected patient has an original CAD-RADS score of 5, therefore it belongs to class 2 in our multi-class experiment. In the figure we reported LAD, LCX and RCA for one of the 8 views available for this patient. These three images actually compose a single input image for our network that was correctly classified as representing a patient of class 2 with a probability of 0.88. For each vessel, we can see on the left the top-3 maximally activated patches in order of importance. The yellow dashed areas highlight the regions of the vessel represented by the patches that have a 80x80 pixels dimension and are extracted from the resized image given in input to the network during the forward pass. On the right of each vessel instead, we can see the map produced by Deep SHAP algorithm for the correctly predicted class (class 2 in this case). The output of the model is influenced by pink pixels in a positive way and by blue pixels in a negative way. The input images are shown as nearly transparent grayscale backings behind each of the explanations. On the right of each map, we can see a zoom of the most relevant pixels (red dashed areas).

\begin{figure*}[ht]
    \centering
    \includegraphics[scale=0.5]{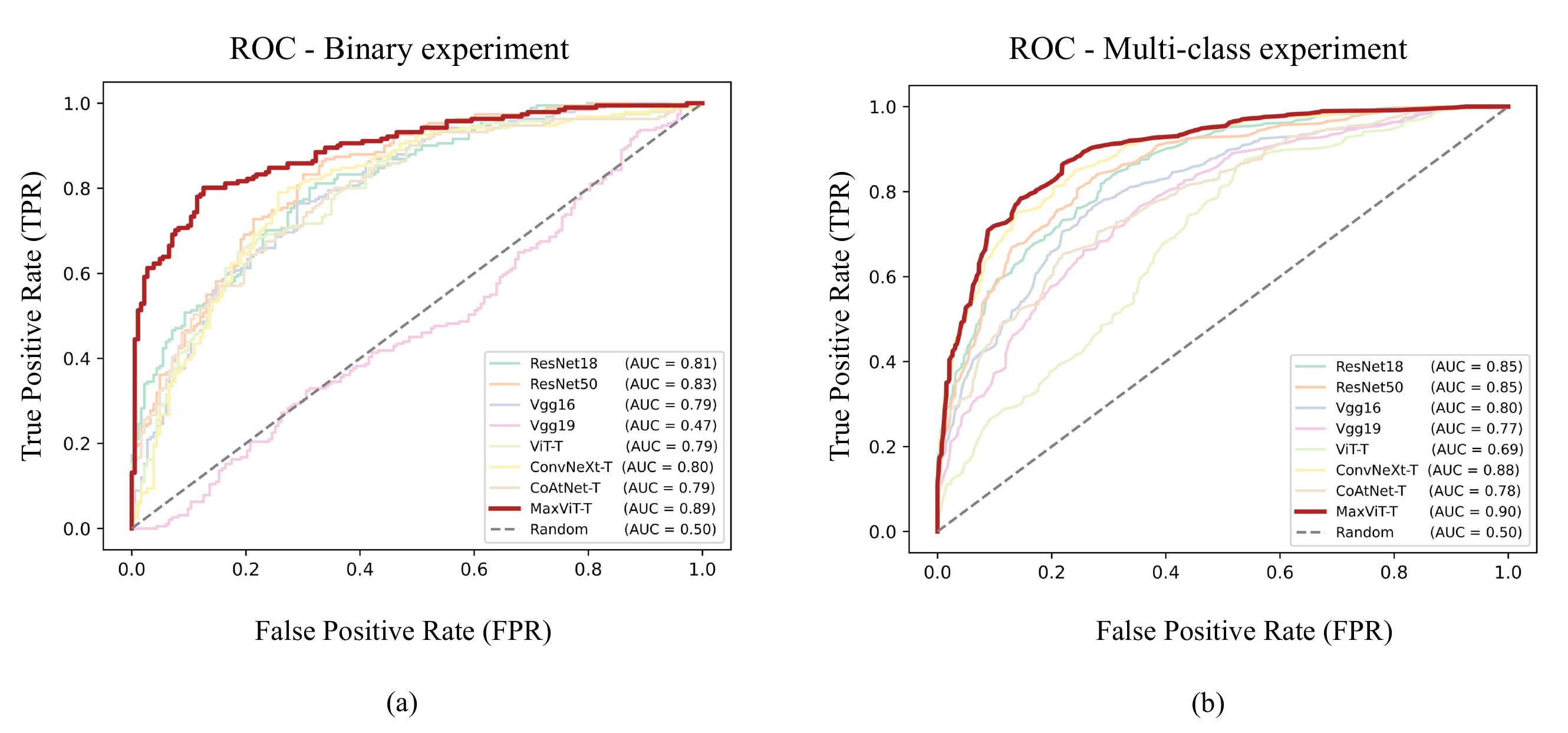}
    \caption{ROC curves of the compared models for the binary (a) and multi-class experiment (b). }
    \label{fig:roc}
\end{figure*}

We also reported in Table \ref{tab:comparison} the resulting per-image AUC and accuracy for both the experiments comparing the performance of the most common convolutional, attention-based and hybrid architectures. DeLong’s test p-values for pairwise AUC comparisons ($\alpha = 0.05$), with the highest AUC as a reference, are provided. For each model the number of parameters and multiply-accumulate operations (MACs) is reported. ROC curves for each experiment and each model are provided in Figure \ref{fig:roc}.

\begin{table}[t]
\caption{Comparison between different architectures: convolutional (ResNet18, ResNet50, Vgg16, Vgg19), transformer (ViT-T) and hybrid (ConvNeXt-T, CAtNet-T, MaxViT-T) models. For each architecture, per-image AUC and Accuracy with 95\% CI are reported. DeLong’s test p-values for pairwise AUC comparisons, with the highest AUC as a reference, are also provided. Finally, in the last two columns we reported the number of parameters and MACs of the compared models.}
\label{tab:comparison}
\resizebox{\linewidth}{!}{
\begin{tabular}{@{}llccccc@{}}
\toprule
\textbf{Model}                     & \textbf{Experiment}  & \multicolumn{1}{l}{\textbf{AUC [95\% CI]}} & \multicolumn{1}{l}{\textbf{Accuracy [95\% CI]}} &  \multicolumn{1}{l}{\textbf{DeLong's p-value}} & \multicolumn{1}{l}{\textbf{Params(M)}} & \multicolumn{1}{l}{\textbf{MACs(G)}} \\ \midrule
\multirow{2}{*}{ResNet18}          & binary               & 0.81 [0.77, 0.85]                         & 0.74 [0.69, 0.79]                              & $<0.001$ & \multirow{2}{*}{11.18}                 & \multirow{2}{*}{1.82}                \\
                                   & multi-class          & 0.85 [0.82, 0.88]                         & 0.82 [0.78, 0.85]  & $<0.001$ &                                                                  &                                      \\
\multicolumn{6}{l}{}                                                                                                                                                                                                                   \\
\multirow{2}{*}{ResNet50}          & binary               & 0.83 [0.79, 0.86]                         & 0.76 [0.71, 0.81]                              & $<0.001$ & \multirow{2}{*}{23.51}                 & \multirow{2}{*}{4.13}                \\
                                   & multi-class          & 0.85 [0.82, 0.89]                         & 0.81 [0.77, 0.84]   & $0.01$                           &                                        &                                      \\
\multicolumn{6}{l}{}                                                                                                                                                                                                                   \\
\multirow{2}{*}{Vgg16}             & binary               & 0.79 [0.75, 0.84]                         & 0.72 [0.68, 0.77]                              & $<0.001$ & \multirow{2}{*}{134.27}                & \multirow{2}{*}{15.47}               \\
                                   & multi-class          & 0.80 [0.76, 0.84]                         & 0.78 [0.75, 0.82]  & $<0.001$                            &                                        &                                      \\
\multicolumn{6}{l}{}                                                                                                                                                                                                                   \\
\multirow{2}{*}{Vgg19}                              & binary               & 0.47 [0.42, 0.54]                         & 0.51 [0.46, 0.56]                              & $<0.005$ & \multirow{2}{*}{139.58}                & \multirow{2}{*}{19.63}               \\
                                   & multi-class          & 0.77 [0.73, 0.81]                         & 0.76 [0.73, 0.80]  & $<0.001$                            &                                        &                                      \\
\multicolumn{6}{l}{}                                                                                                                                                                                                                   \\
\multirow{2}{*}{ViT-T}             & binary               & 0.79 [0.74, 0.83]                         & 0.72 [0.68, 0.76]                              & $<0.001$ & \multirow{2}{*}{5.49}                  & \multirow{2}{*}{1.08}                \\
                                   & multi-class          & 0.69 [0.65, 0.73]                         & 0.71 [0.67, 0.74]   & $<0.001$                           &                                        &                                      \\
\multicolumn{6}{l}{}                                                                                                                                                                                                                   \\
\multirow{2}{*}{ConvNeXt-T}        & binary               & 0.80 [0.75, 0.84]                         & 0.76 [0.72, 0.81]                              & $<0.001$ & \multirow{2}{*}{27.80}                 & \multirow{2}{*}{4.45}                \\
                                   & multi-class          & 0.88 [0.85, 0.91]                         & 0.83 [0.80, 0.86]  & 0.44                            &                                        &                                      \\
\multicolumn{6}{l}{}                                                                                                                                                                                                                   \\
\multirow{2}{*}{CoAtNet-T}         & binary               & 0.79 [0.75, 0.83]                         & 0.72 [0.75, 0.83]                               & $<0.001$ & \multirow{2}{*}{5.48}                  & \multirow{2}{*}{7.64}                \\
                                   & multi-class          & 0.78 [0.74, 0.82]                         & 0.77 [0.74, 0.81]    & $<0.001$                          &                                        &                                      \\
\multicolumn{6}{l}{}                                                                                                                                                                                                                   \\
\multirow{2}{*}{\textbf{MaxViT-T}} & \textbf{binary}      & \textbf{0.89 [0.86, 0.93]}                & \textbf{0.82 [0.78, 0.86]}                     & \multirow{2}{*}{-} & \multirow{2}{*}{28.45}                 & \multirow{2}{*}{4.89}                \\
                                   & \textbf{multi-class} & \textbf{0.90 [0.86, 0.92]}                & \textbf{0.86 [0.83, 0.89]}                     &                                        &                                      \\ \bottomrule
\end{tabular}}
\end{table}

\section{Discussion}
\label{sec:discussion}
The proposed pipeline achieved high performance in all the reported metrics for both the binary and multi-class experiment, showing a per-patient AUC [95\% CI] of 0.87 [0.76, 0.95] and 0.93 [0.84, 0.99], respectively. To the best of our knowledge, this is the first work classifying the sequence of the three main coronary arteries relying only on patient-wise CAD-RADS scores during the training procedure. None of the previous works are directly comparable to our method as most of them rely on vessel, segment, lesion-wise annotations. We were able to achieve highly accurate results without the need of any additional annotation step. This demonstrates that it is possible to train a model to assign an overall score to a patient's three main coronary arteries without requiring additional clinical annotations, which are not part of the screening routine. The work conceptually most similar to ours is that presented by \cite{li2022automatic}. Although they used a completely different methodological approach, their aim was similar to ours since they trained a model to assign patient-wise CAD-RADS scores based on the whole coronary tree obtaining an AUC of 0.737 for a binary classification task. 
Compared with works that used finer annotation steps, we might expect our performance to be poorer due to the smaller amount of information used during the learning procedure. Nonetheless, we achieved very high performance in both tasks, often outperforming previous works that used finer annotations. It must be noted, however, that direct comparison is challenging as the datasets used are not publicly available. The strength of our work primarily relies on the network's ability to autonomously learn complex features, while taking the entire context into account, rather than considering single vessels, segments, or lesions.

From the results shown in Table \ref{tab:comparison}, it can be seen that the chosen architecture clearly outperformed all the other methods included in the comparison. In particular, Vgg19, which is the deepest fully convolutional architecture, and ViT-T which is the only vision transformer model, were the worst performing algorithms, showing poor generalization capabilities in both the experiments. 

For the binary experiment, we can recognize two areas from the t-SNE plots in Figure \ref{fig:tsne} (a,b) with the patients in the lower left corner representing subjects with higher CAD-RADS score. For the multi-class experiment (c,d), instead, we can recognize three main groups basically representing patients with zero, mild and severe stenosis. It is interesting to notice that the separation between healthy subjects (green) and patients with high CAD-RADS scores (red) is clear-cut. Interestingly, most of the errors occurred in the intermediate group. This finding reflects the fact that misclassifications in the test set were always off by only one class (e.g., a patient classified as belonging to class 1 when he was actually in class 0, but never misclassified as belonging to class 2). Furthermore, when we examine the plot colored by the original CAD-RADS scores (d), we can observe that misclassifications in class 2 only occurred in patients with CAD-RADS scores lower than 5. This finding is particularly interesting because t-SNE is an unsupervised technique, yet the distribution of patients in the 2D space reflects the expected clusters based on domain knowledge. 

In Figure \ref{fig:shap}, we can observe both maximally activated patches and higher absolute SHAP values in the most critical regions of the vessels. Regarding the LAD vessel, we can see that most of the pixels are blue in the central part, indicating that, based on those pixels, the predicted class could have been lower than 2. In the case of the LCX vessel, there are several clearly visible occlusions in the final part of the artery, and the SHAP values are higher in absolute value in this area, with pink pixels pushing the prediction towards the worst CAD class, while blue pixels attenuate this effect due to partially normal areas of the vessel. For the RCA, we can see two separate areas where the pixels are mostly blue and mostly pink, respectively. In the region occupied by the pink pixels, we can observe a slightly brighter area in the original image, which undoubtedly influenced the model output positively in predicting the higher CAD class. However, the final class is assigned based on all three coronary vessels, that represents 3 channels of a single input image and, in this case, it led to the correct prediction of the patient's outcome phenotype (class 2, i.e. CAD-RADS = 4-5).

This approach could be helpful in visually inspecting which part of the image the model gives more importance to, and to check for potential artefacts that may have erroneously influenced the final classification. This tool could be beneficial in better examining suspicious cases to quickly detect possible biases that could indicate that the algorithm may not be trustworthy in those cases. These user-friendly maps offer to non-expert clinical users a transparent way to assess the reliability of the prediction, avoiding a completely black-box approach.

The strength of this work relies on the pipeline ability to achieve highly accurate CAD-RADS predictions without the need for any additional effort in the clinical practice, such as annotating single vessels, segments, or lesions. This could be advantageous in a real clinical setting where any additional step, not included in the standard clinical routine, would be limited to specific research studies. Furthermore, in our pipeline, expert clinical user manual intervention is minimized, thereby avoiding several time-consuming and operator-dependent steps. Another crucial aspect is the way we composed the input data for the network. By always using the three main coronary arteries together as a single input, we trained a network to associate the CAD-RADS score to a sequence of arteries rather than each single vessel, exactly as a radiologist would do when assigning a patient-wise CAD-RADS score.

This work has some limitations that need to be addressed. First, the overall dataset, although significant for a clinical task, is relatively small. To mitigate this issue, we used a SOTA architecture specifically designed to work under all data regimes, and employed several technical strategies, including pre-training, cross-validation, data augmentation, weight and LR decay and label-smoothing, to avoid overfitting. However, the performance of the proposed algorithms could be possibly further improved by training on a larger dataset. Additionally, using data from a single clinical center limits the generalization ability of our results. It is essential to test the models on independent populations to assess their generalization capabilities before application in clinical practice. 
Although during screening procedures the goal is more to quickly identify macro-categories of patients needing further assessment or to rule-out healthy subjects, a larger dataset could allow testing a multi-class classification approach that takes into account all CAD-RADS scores separately. Moreover, to restrict our analysis, in this study we did not include modifiers to describe patients with stents (modifier S), vulnerable plaque features (modifier V), or grafts (modifier G). From a clinical point of view, it would be interesting to evaluate the generalizability of the models to these sub-categories as well. 
From a technical perspective, possible future extensions of this work could be substituting the proposed imputing strategy with a deep generative algorithm, specifically trained to generate synthetic samples representing healthy vessels. Although this approach would add a layer of complexity requiring substantial computational resources, it would be interesting to explore how the models' performance would be influenced by this choice.

\section{Conclusions}
\label{sec:conclusions}
We proposed a fully automated pipeline able to classify images representing 2D longitudinal cross-sections extracted from CCTA scans of the three main coronary arteries, based on a fine-tuned Multi-Axis Vision Transformer model. The highly accurate results obtained in the two explored tasks (identify patients in need for further investigations and classify the patients according to the severity of the occlusion) suggest the great potential of the proposed approach. This is the first work that does not require any additional annotation step, which is not part of the clinical routine, and uses instead, a learning procedure that perfectly emulates the clinical screening process. Such a tool would be a useful decision support system in the clinical practice, able to help the radiologists in quickly identify severe patients as well as completely healthy subjects and better inspect borderline cases with the help of intuitive and visually explainable methods.

\section*{Declaration of Competing Interests}
Riccardo Bellazzi is co-founder and shareholder of two spin-offs of the University of Pavia (Engenome s.r.l. and Biomeris s.r.l.) that operate in the field of data management, artificial intelligence and bioinformatics.

\section*{Funding}
This work was supported by Fondazione Regionale per la Ricerca Biomedica (FRRB), Research Grant no. CP2\_14/2018, Project "INTESTRAT-CAD" (PI: Gualtiero I. Colombo).\\

The study protocol conformed to the principle of the Declaration of Helsinki and was approved by the "Ethics Committee of the IRCCS Istituto Europeo di Oncologia and Centro Cardiologico Monzino" (protocol code R329/15 - CCM 341, date of approval 23/09/2015). All recruited patients signed the written informed consent and participants did agree to share their de-identified information.

\bibliographystyle{model2-names.bst}\biboptions{authoryear}
\bibliography{manuscript}

\end{document}